\documentclass[prd, twocolumn, nofootinbib, floatfix]{revtex4-1}

\usepackage{amsmath}
\usepackage{graphicx}
\usepackage{dcolumn}
\usepackage{bm}
\usepackage{epsfig}
\usepackage{amssymb,latexsym,mathrsfs}
\usepackage{graphicx}
\usepackage{color}
\usepackage{hyperref}

\hypersetup{
    colorlinks=true,
    linkcolor=red,
    citecolor=blue,
} 

\newcommand{\be}{\begin{equation}}
\newcommand{\ee}{\end{equation}}
\newcommand{\bs}{\begin{split}} 
\newcommand{\bea}{\begin{eqnarray}}
\newcommand{\eea}{\end{eqnarray}}

\newcommand{\om}{\Omega_m}

\newcommand{\fs}{f\sigma_8}

\begin{document}

\title{Cosmic Growth and Expansion Conjoined} 
\author{Eric V.\ Linder${}^{1,2}$} 
\affiliation{${}^1$Berkeley Center for Cosmological Physics \& Berkeley Lab, 
University of California, Berkeley, CA 94720, USA\\ 
${}^2$Energetic Cosmos Laboratory, Nazarbayev University, Astana, 
Kazakhstan 010000} 

\begin{abstract}
Cosmological measurements of both the expansion history and growth 
history have matured, and the two together provide an important test 
of general relativity. We consider their joint evolutionary track, 
showing that this has advantages in distinguishing cosmologies relative to 
considering them individually or at isolated redshifts. In particular, 
the joint comparison relaxes the shape degeneracy that makes $f\sigma_8(z)$ 
curves difficult to separate from the overall growth amplitude. The 
conjoined method further helps visualization of which combinations of 
redshift ranges provide the clearest discrimination. We examine standard 
dark energy cosmologies, modified gravity, and ``stuttering'' growth, 
each showing distinct signatures. 
\end{abstract} 

\date{\today} 

\maketitle

\section{Introduction} 

The histories of the expansion of the universe and the growth of large 
scale structures within it are key observables that provide insights into 
the cosmological model. In particular, within general relativity the two 
are tightly tied together, even more so within models close to the 
concordance cosmological constant plus cold dark matter, $\Lambda$CDM. 
In standard models, these histories are quite smooth and gently varying, 
generally on a Hubble expansion timescale, and thus discriminating between 
cosmologies is not easy, even with reasonably precise measurements. That is, 
one does not have sharp or oscillating features the way one does when 
analyzing, for example, cosmic microwave background (CMB) power spectra. 

Although the amplitudes of the growth or growth rate history, for example, 
may differ between cosmologies, this is often nearly degenerate with the 
initial conditions of the mass fluctuations, or the present mass fluctuation 
amplitude $\sigma_8$. Without a clear distinction in the shape of the history 
curve over the epochs where precise data exists, this makes cosmological 
characterization problematic. 

We therefore seek a way to interpret the data such that the difference in 
the shapes 
of the evolutionary tracks becomes more pronounced. The simple solution we 
find is to contrast the expansion history in terms of the Hubble expansion 
rate $H(z)$ directly with the growth history in terms of the growth rate 
$\fs(z)$, rather than each as a function of redshift. 
Just as the tracks in a Hertzsprung-Russell (HR) diagram of luminosity vs 
temperature, or a supernova plot of magnitude vs color, can illuminate the 
physics more clearly than plotting either vs its dependent variable (age 
or time), so too do the cosmic histories when plotted 
against each other rather than as a function of time or redshift. Of course 
no extra physics is added by this change in visualization, just readier 
recognition, identification, and interpretation of the existing physics, 
i.e.\ deviations from general relativity or standard cosmology and 
particular redshift ranges of interest. 

In Sec.~\ref{sec:prev} we exhibit the difficulties in distinguishing 
cosmologies in the standard approach of the evolutionary tracks vs time, 
as well as a new combination of both histories. We introduce the HR-type 
approach of considering histories conjointly in Sec.~\ref{sec:tog} 
and investigate it for several types of cosmologies, including quintessence, 
modified gravity, and stuttering growth. Section~\ref{sec:data} discusses 
the impact of future measurements on exploring the cosmic expansion and 
growth histories, and we conclude in Sec.~\ref{sec:concl}.

\section{Histories vs Time} \label{sec:prev} 

The expansion history can be most directly considered in terms 
of the Hubble expansion rate $H(z)=\dot a/a$, where $a=1/(1+z)$ 
is the cosmic scale factor and $z$ is the redshift. The Hubble 
parameter sets the scale for cosmic distances and age. Distances are 
integrals over the expansion history so $H(z)$ is more 
incisive concerning the conditions at a given redshift. The 
growth history similarly is best examined by means of an 
instantaneous quantity, the growth rate $f=d\ln D/d\ln a$, where 
$D(z)$ is the overall growth factor from some initial condition 
to a redshift $z$. 

In fact, observations are sensitive to a product $\fs\propto fD$, 
where $\sigma_8(z)$ measures the rms mass fluctuation amplitude 
at redshift $z$. Remarkably, measurements of the clustering of 
large scale structure provide both $H$ and $\fs$, so a cosmic 
redshift survey delivers both the expansion history and growth 
history. The expansion rate comes from using the baryon 
acoustic oscillations in the clustering pattern as a standard 
ruler, and in particular the radial distances measure 
$H(z)\,r_d$, where $r_d$ is the sound horizon at the baryon drag 
epoch in the early universe. The growth rate comes from 
redshift space distortions of the clustering pattern, caused by 
the (gravitationally induced) velocities of the galaxies or other 
tracers. 

\begin{figure}[htbp!]
\includegraphics[width=\columnwidth]{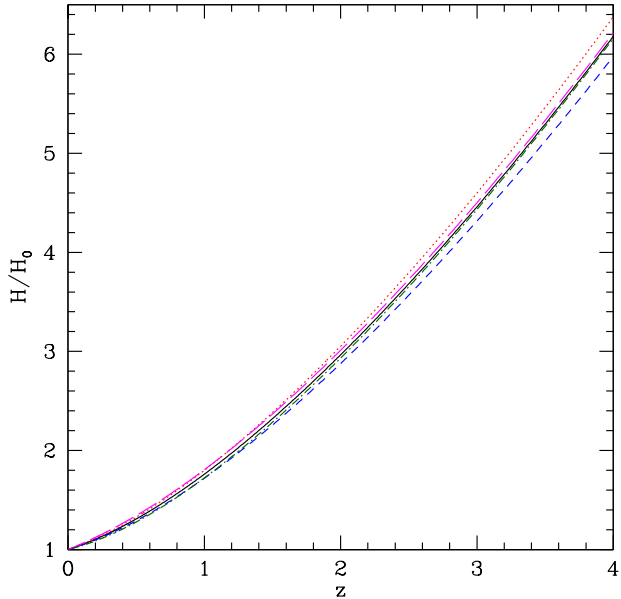} 
\caption{
The Hubble expansion parameter $H(z)/H_0$ is plotted as a function of 
redshift for 
cosmological models with matter density $\om=0.28$, 0.3, 0.32 (blue dashed, 
black solid, red dotted curves respectively) for LCDM 
($w=-1$), plus $w=-0.9$, $-1.1$ (magenta long dashed and green dot-dashed 
curves respectively) for $\om=0.3$. The curves are smooth without localized 
features and degeneracies are apparent. 
} 
\label{fig:hz} 
\end{figure}

\begin{figure}[htbp!]
\includegraphics[width=\columnwidth]{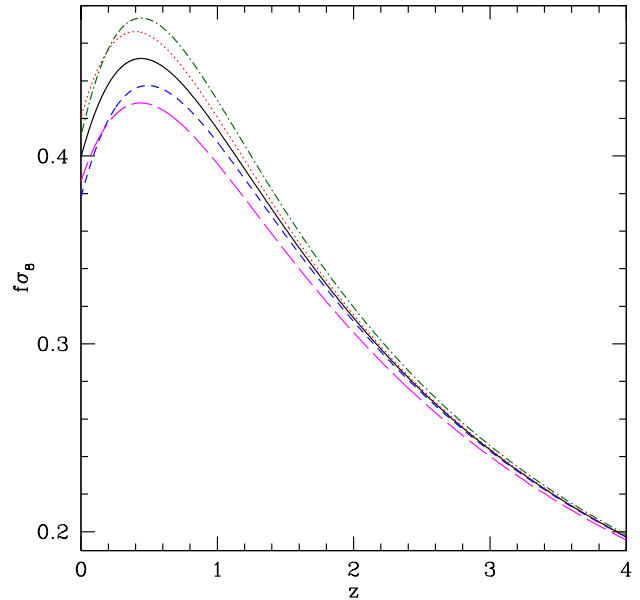} 
\caption{
As Fig.~\ref{fig:hz} but plotting the growth rate $\fs$ as a function of 
redshift. The models show 
more discrimination but at $z\approx0.5$--1.5 are mostly similar shapes 
with scaled amplitudes. 
} 
\label{fig:fsz} 
\end{figure}

Both $H(z)$ and $\fs(z)$ tend to be smooth, slowly varying 
functions. In particular, $H(z)$ is generally monotonic and 
changes on a Hubble, or e-folding, time scale, while $\fs(z)$ 
has a broad peak which means that its value changes little 
during recent cosmic history where the data is best measured. 
Figure~\ref{fig:hz} shows $H(z)$ for several models while 
Fig.~\ref{fig:fsz} illustrates the properties of $\fs(z)$ for the 
same models.

Note that not only is $\fs(z)$ reasonably constant, but 
the shapes of the curves for different cosmologies are fairly 
similar, mostly being simply offsets in amplitude. That is, they 
have different rms mass fluctuation amplitudes at present, 
$\sigma_8$, but otherwise look similar for 
different cosmological physics. Standard models will have this 
broad peak due to simple physical constraints: at high redshift, 
in the matter dominated era, $f\to1$ and $D\to a$ so 
$\fs\propto a$ independent of model specifics. The linear increase with 
$a$ in $\fs$ is counteracted at recent times by the 
suppression of growth caused by accelerating expansion -- this 
reduces $f$ below 1 and causes $D$ to grow more slowly so the net 
effect is a gradual turnover during the accelerating epoch. 

This lack of strong, cosmology dependent features in $H$ and 
$\fs$ is disappointing since recent redshift surveys such as 
BOSS \cite{160703155,161003506,160903697,160703153,160703150,160703149,160703148,160703147,160703145,160703143,160600439}
and WiggleZ \cite{13025178,12102130,12043674} 
have demonstrated precision measurements of 
these quantities and next generation spectroscopic surveys such 
as PFS, DESI, Euclid, and WFIRST will greatly improve on this. 
While a full statistical analysis will constrain the cosmological 
parameters, the visual ``smoking gun'' of a deviation from the 
standard model may not be apparent, and the results may depend 
more on the parametrization and the build up of signal to noise 
over redshift range. Thus, the motivation exists to find a more 
incisive, ideally visually clear method of using this expansion 
and growth data to discriminate between cosmologies. 

In general relativity, expansion and growth are tied together, 
with expansion (and any microphysics such as sound speed) 
determining growth. That is, the linear growth equation depends 
solely on $H(z)$ and the present matter density. However, 
cosmologies where accelerated expansion is caused by extensions 
to Einstein gravity generally break this close relation, allowing 
for greater changes to the $H$ and $\fs$ histories. This suggests 
that simultaneous consideration of these two functions may give 
greater insight. This has been explored, with some interesting 
results, at individual redshifts, i.e.\ the expansion at redshift 
$z_1$ vs the growth at redshift $z_1$ 
\cite{song1,song2,song3,dr11a,dr11b,160703155}. 

Here we extend this to conjoint investigation of expansion and 
growth as full functions, i.e.\ their histories or evolutionary 
tracks. The first thing one might try, motivated by the above 
discussion about the generic behavior of $\fs$ (and $H$) in the 
matter dominated era, is to combine the functions together. 
We have good reasons to believe that a matter dominated era must exist, 
whatever the late time cosmology: breaking matter 
domination would give a huge Sachs-Wolfe effect on the CMB in 
contradiction to observations, plus severely impact the formation 
of large scale structure. Figure~1 of \cite{dbl} shows the 
dramatic effect of even 0.1 e-fold of early acceleration on the 
CMB. 

\begin{figure}[htbp!]
\includegraphics[width=\columnwidth]{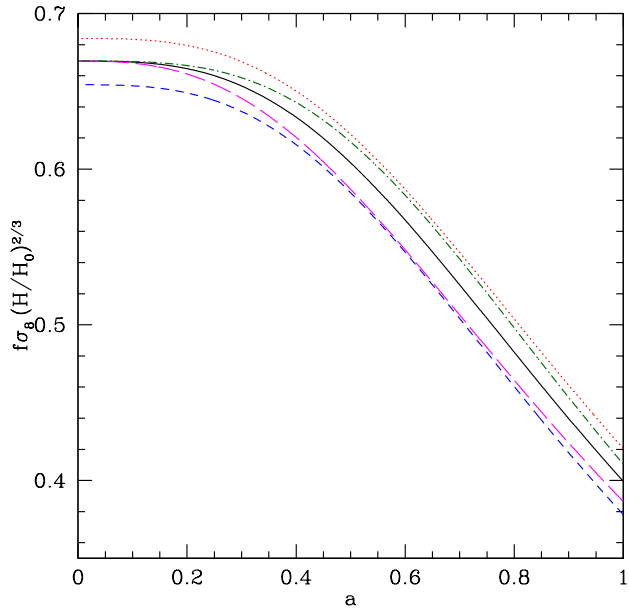} 
\caption{
The combination $\fs(H/H_0)^{2/3}$ is plotted vs scale factor for the 
models in Fig.~\ref{fig:hz}. This combination goes to a constant at high 
redshift (small $a$). This constancy provides an important test of early 
matter domination and general relativity, but is at redshifts $z\gtrsim4$ 
where observations are difficult. At redshifts $z\lesssim1.5$, the 
degeneracy between $\om$ and $w$ is apparent, so measurements out to 
$z\approx3$ are important. 
} 
\label{fig:fsh23} 
\end{figure}

Given early matter domination, recall that $\fs\propto a$ and 
$H^2\propto a^{-3}$. This suggests that all reasonable 
cosmologies should go to $\fs H^{2/3}=\ $constant during that 
epoch. To investigate whether convolving the expansion and 
growth histories in such a way improves distinction between 
models, we plot this combination vs redshift in 
Fig.~\ref{fig:fsh23}.

The evolutionary tracks are quite similar, lying in a fairly 
narrow band. In particular, the curve shapes do not have any 
distinguishing features. The amplitudes go to a constant 
involving $(\om h^2)^{2/3}A_s^{1/2}$ at high redshift, where 
$A_s$ is the scalar perturbation power amplitude of the CMB. 
Since the CMB similarly tightly constrains the combination 
$\om h^2$, models start out at small $a$ 
close together and are mostly affected by more recent growth 
effects such as suppression from cosmic acceleration. Because of 
the lack of shape features, the $\fs H^{2/3}$ test seems better 
suited as a consistency test, though precision data during the 
high redshift ($z\gtrsim 3$) constant regime is difficult to 
obtain. 

The approach in the next section appears much more promising to 
discriminate visually between cosmologies.

\section{History vs History} \label{sec:tog} 

Another approach to conjoint analysis of the expansion and growth 
histories is, rather than combining the functions, to contrast 
them. That is, consider the histories as a function of each other, 
rather than of time for either the individual or convolved histories. 
The time dependence will run along the curves in the two dimensional 
space of expansion vs growth, or more specifically $H$ vs $\fs$. 

\begin{figure}[htbp!]
\includegraphics[width=\columnwidth]{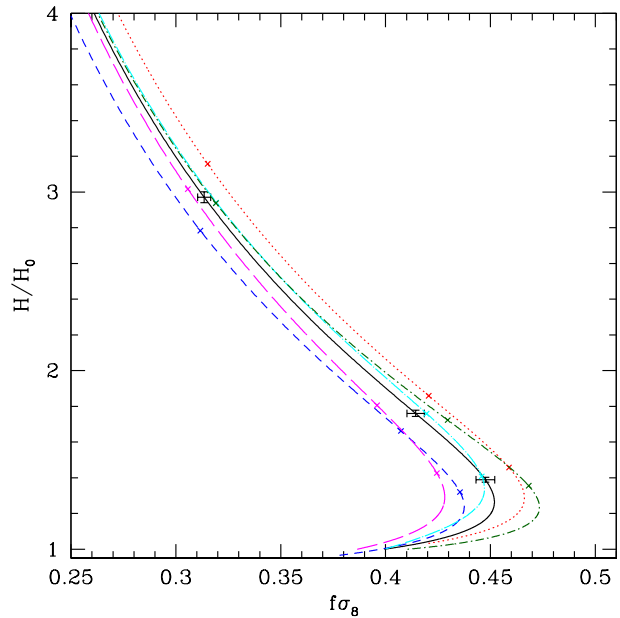} 
\caption{
The expansion rate $H(z)/H_0$ and growth rate $\fs$ are plotted 
against each other for the cosmological models in Fig.~\ref{fig:hz}. 
Redshift, or scale factor, runs along the curves; small x's 
indicate $z=0.6$, 1, 2 along the individual curves (from bottom to top). 
The cyan, 
dot-long dash curve is a mirage dark energy model. Plotted jointly, 
the expansion history and growth history exhibit more definite 
curvature, and discrimination beyond $z\gtrsim1.5$. Error bars 
along the $\om=0.3$ LCDM (solid black) curve indicate 1\% 
uncertainties in measurements of expansion and of growth at 
$z=0.6$, 1, 2. 
} 
\label{fig:hvsfs} 
\end{figure}

Figure~\ref{fig:hvsfs} shows that this has useful 
characteristics, including a fairly well defined bump or wiggle 
that could clarify visually distinctions between cosmological models. 
Moreover, the interpretation in terms of physics remains in the 
forefront: e.g.\ for a given expansion rate (horizontal cut), the 
growth rate is seen to be 
enhanced or suppressed relative to a fiducial model. Because the 
axes involve rates, this gives a focused view rather than being 
complicated by inertia from earlier or later conditions, as the 
overall growth factor or distances would be.

The models have been normalized to all have the same 
$\om h^2$, which is well measured by the CMB. Thus at high 
redshift the tracks are all parallel to each other. While the 
y-axis is labeled as $H/H_0$, this is actually the $H_0$ of the 
$\om=0.3$ model, so models with different $\om$ have 
$H(z=0)/H_0=(0.3/\om)^{1/2}$ rather than unity. At intermediate 
redshifts, the history vs history tracks show a prominent 
wiggle in the range $z\approx 0.4-1$. Since such a feature aids 
distinction between models, this approach is particularly useful 
since this region has the most accurate measurements with 
current, and much future, data. 

Shifts in the dark energy equation of state do not behave 
substantially 
differently from shifts in the matter density. This holds as 
well for a time evolving equation of state; the mirage dark energy 
model with strong evolution $(w_0,w_a)=(-0.8,-0.732)$ that 
nevertheless matches the CMB distance to last scattering of the 
$\Lambda$CDM model \cite{07080024} lies within the envelope defined 
by the $w=-0.9$ and $w=-1.1$ curves over the redshift range plotted. 
This means that most viable standard models, i.e.\ within 
general relativity and not too far from $\Lambda$CDM, have 
roughly the same shape and lie within a band around the fiducial 
model. They mostly differ in the exact degree and location of 
the wiggle. 

\begin{figure}[htbp!]
\includegraphics[width=\columnwidth]{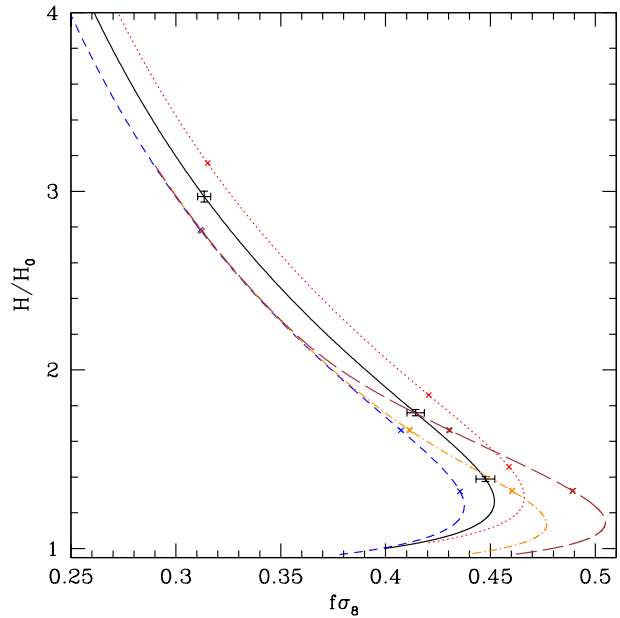} 
\caption{
As Fig.~\ref{fig:hvsfs} but showing a modified gravity model instead 
of the dark energy $w$ curves. The modified gravity curves are for an 
exponential $f(R)$ gravity model and exhibit scale dependence in 
growth, with a clear difference between wave numbers 
$k=0.02\,h$/Mpc (orange, dot-dash curve) and $k=0.1\,h$/Mpc 
(brown, long dashed curve). The $f(R)$ model has an expansion 
history equivalent to $\om=0.28$ LCDM (blue, short dashed curve), 
but a very different growth history. 
} 
\label{fig:hvsfsmod} 
\end{figure}

Next we consider models that enhance or suppress growth relative 
to standard models within general relativity. First we examine 
a modified gravity model, the exponential $f(R)$ gravity of 
\cite{expfr} with $c=3$. This has an expansion history extremely 
close to that of $\Lambda$CDM but has the generically enhanced 
growth of scalar-tensor 
gravity. Figure~\ref{fig:hvsfsmod} illustrates a strong effect 
on the wiggle feature, with an enhanced growth rate at constant 
expansion rate (note the x's of the modified gravity model are 
horizontally aligned with the $\Lambda$CDM model with the same 
matter density). Note that the wiggle would veer even further to 
the right if we showed this model with the fiducial $\om=0.3$ 
rather than $\om=0.28$ as used. That is, already with $\om=0.28$ 
the $f(R)$ track and its wiggle sweep all the way from the 
leftmost standard curve past the rightmost one. Thus, such a 
modified gravity model should be clearly distinguishable -- 
especially with low redshift measurements.

Modified gravity models in general have scale dependent growth 
as well. Figure~\ref{fig:hvsfsmod} illustrates this by showing 
the history vs history tracks for two different values of the 
density perturbation wave number, $k=0.02$ and $0.1\,h$/Mpc. 
At high redshift, the modified gravity theory acts like general 
relativity but deviates more recently in the low Ricci scalar 
curvature regime. On larger scales (smaller $k$) the deviation 
in growth at low redshift is more mild as scalar-tensor theories 
generally involve $k^2$ corrections to Einstein gravity. This 
gives the characteristic enhanced, late, scale dependent wiggle 
seen in Fig.~\ref{fig:hvsfsmod}. Note that the expansion 
histories remain identical, to each other and to $\om=0.28$ 
$\Lambda$CDM, as shown by the x's lined up horizontally at the 
same values of $H(z)/H_0$. Thus, signatures of modified gravity 
are particularly visible in this expansion history vs growth 
history plane. 

The opposite case of suppressed growth is more difficult to 
achieve within modified gravity for the standard expansion history 
(since scalar-tensor theories generate additional attractive 
forces that enhance growth). Instead, we use the superdecelerating 
dark energy model of \cite{10064632,dbl}. This is purely 
phenomenological and 
involves a period of enhanced dark energy density with $w=-1$ at 
early times (though still too little to affect significantly the 
CMB or cause an epoch of early acceleration) then a step up to 
the maximal equation of state $w=+1$ for a 
standard scalar field, in order to dilute quickly the extra 
density (the superdeceleration), and a restoration to $w=-1$ at 
more recent times. The enhanced dark energy density reduces the 
source term in the growth equation and suppresses growth. 
This causes a ``stutter'' in growth, where basically the momentum 
of the growth ($\dot\delta$) drops significantly  as matter 
domination wanes and the enhanced dark energy density comes into 
play. Even though 
the conditions after the step back down to $w=-1$ are identical 
to low redshift $\Lambda$CDM, the growth has been stunted and 
takes time to recover. 

\begin{figure}[htbp!]
\includegraphics[width=\columnwidth]{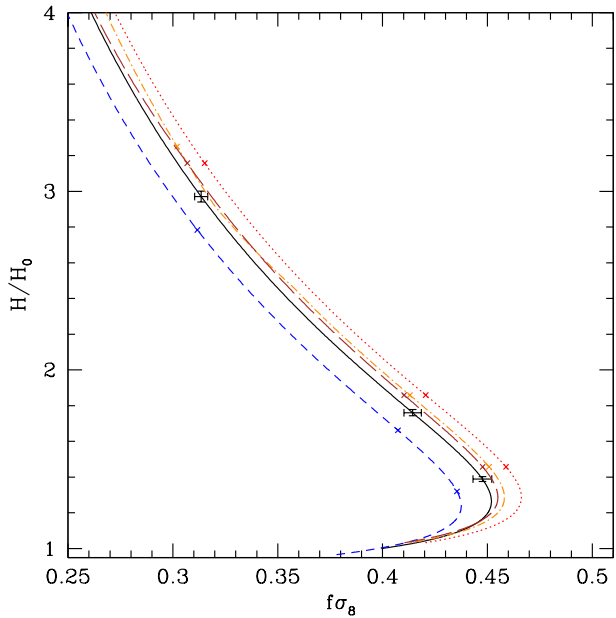} 
\caption{
As Fig.~\ref{fig:hvsfs} but showing a suppressed growth superdecelerating 
model instead of the dark energy $w$ curves. The stunted growth curves 
show results from $N=0.2$ $e$-folds of superdeceleration at $z_d=2$ (brown, 
long dashed curve) and $N=0.1$ at $z_d=1.5$ (orange, dot-dash curve), 
both in accord with CMB distance to 
last scattering measurements and having $\om=0.32$. At redshifts below  
the transition redshift, the expansion history is identical to that of 
$\om=0.32$ LCDM and the expansion vs growth curves gradually become 
horizontal offsets of that model, as the growth rate recovers at a 
suppressed growth amplitude. 
} 
\label{fig:hvsfsbox} 
\end{figure}

The results in Fig.~\ref{fig:hvsfsbox} show the impact of this 
model. We choose a step of length $N=0.2$ $e$-folds ending at 
$z_d=2$, falling in the middle of the allowed region of Fig.~5 in 
\cite{10064632}. Too long a period of superdeceleration, at 
too recent an epoch, changes the CMB distance to last scattering 
and causes an excessive integrated Sachs-Wolfe effect (see 
\cite{10064632,dbl} for detailed discussion of the physical 
effects). Growth is indeed suppressed for nearly the same 
expansion history over the range of redshifts plotted. The wiggle 
now is pushed to the left, from the rightmost standard curve 
(we use this model with $\om=0.32$) toward those with lower 
matter density.

To understand the behavior, note that at high redshift, well above 
the transition in behavior, there is enhanced dark energy density 
(by a factor $e^{6N}$). 
During the matter dominated regime, this has little effect and the 
superdecelerating models used join the standard $\om=0.32$ curve. 
However, the enhanced dark energy density (and hence lower matter 
density for a flat universe) has more of an impact on expansion 
and growth at higher redshift than the standard model, increasing 
the expansion rate but slowing the growth rate, and the conjoint 
history curve begins to bend to the left, acting as reduced 
matter density curves in both expansion and growth. Indeed, if 
a period of early dark energy domination were allowed to occur, then 
the growth rate would approach zero. However, such a strong effect 
would leave clear signatures in the CMB (see \cite{10064632,dbl}) 
and is ruled out by current measurements. 

During the $e$-folds of 
superdeceleration, the dark energy density drops; to preserve the 
same value of dark energy density today, a larger number 
$N$ of e-folds implies higher early dark energy density, so $N$ 
is tightly constrained. After the superdeceleration period ends, 
the expansion history is identical to that of the corresponding 
matter density model, and the growth rate can recover from its 
``stutter'' where the momentum of growth $\dot\delta$ was 
suppressed, but the amplitude of growth $\sigma_8$ is at a lower 
level. Overall this shifts the conjoint history tracks of these 
models to the left, roughly parallel to the standard cosmology 
track. This can lead to degeneracy with lower matter density 
standard cosmologies (e.g.\ $\om\approx0.31$ for the case shown 
in Fig.~\ref{fig:hvsfsbox}) -- unless the histories are measured out 
to sufficiently high redshift near the transition, or 
accurately at low redshift $z\lesssim0.5$ below the wiggle 
where different standard cosmology tracks are not parallel. 

Thus, the conjoint histories diagram appears to be an effective 
method for understanding the behavior of expansion and growth 
viewed together, and distinguishing models both within standard 
cosmology, and especially those that deviate from it by enhancing 
or suppressing growth, e.g.\ through modified gravity or a 
stutter in sourcing growth. Moreover, it highlights the different 
and mutually supportive roles of expansion and growth measurements 
near the wiggle ($z\approx0.5$), at low redshift $z\lesssim0.5$, 
and at high redshift $z\approx2$--3 to identify the cosmological 
physics.

\section{Future Constraints} \label{sec:data} 

Future spectroscopic surveys will use galaxy and quasar 
clustering to measure the expansion rate $H$ and growth rate 
$\fs$ with subpercent precision over a wide range of redshift. 
For example, DESI will extend up to $z\approx1.5$ with galaxies 
and can reach higher redshift with quasars 
and the Lyman-$\alpha$ forest, while Euclid 
and WFIRST will cover $z\approx 1-2.5$ at such precision 
\cite{13084164,13095380}. We have indicated in Figs.~\ref{fig:hvsfs}, 
\ref{fig:hvsfsmod}, and \ref{fig:hvsfsbox} the $1\sigma$ 
constraints assuming 1\% measurements 
of each quantity at various redshifts (treating them 
approximately as uncorrelated since $H$ arises from radial BAO 
and $\fs$ from redshift space distortions, using different scales 
in the data; the qualitative conclusions do not depend strongly 
on this). 

Obtaining highly accurate growth measurements at $z\gtrsim3$ 
will be challenging, but useful for breaking degeneracies 
between models. An intriguing area for further concentration 
is the low redshift regime, $z\lesssim0.5$, where the wiggle 
peaks and then the conjoined history curves change shape. 
Surveys are beginning to map substantial fractions of the 
available volume here, and peculiar velocity surveys such as 
the TAIPAN survey or perhaps future supernova surveys may map 
the growth rate \cite{160908247,160804446,14043799}, although 
the restricted volume limits the attainable precision.

\section{Conclusions} \label{sec:concl} 

The cosmic expansion and cosmic growth histories are fundamental 
observables describing our universe. Current surveys are taking the 
first steps at mapping these over substantial parts of the recent 
history, and future surveys will greatly expand this in range and 
accuracy. While these two quantities, $H(z)$ and $\fs(z)$, as a 
function of redshift contain the cosmological information, viewing 
them simultaneously as a conjoined constraint can illuminate important 
aspects of the overall cosmic evolution, especially for cases that 
go beyond standard models within general relativity. 

The conjoined approach can aid in interpretation, e.g.\ clearly 
recognizing enhanced or suppressed growth for a given expansion 
history, or identifying particular redshift ranges of interest. 
The swing, or wiggle, in the conjoined diagram highlights a time 
of greater sensitivity to the cosmological model, and the regions 
of high and low redshift can also identify and point to methods of 
breaking covariances between parameters. 

We exhibited the joint history tracks for four models: a standard 
constant dark energy equation of state, a time varying dark energy 
equation of state that matches the CMB distance to last scattering, 
a modified gravity model than enhances growth -- in a scale dependent 
manner, and a stuttering, 
superdecelerating model that suppresses growth. We discussed the 
physical effects and sensitive redshift range of each, with some 
altering both expansions and growth histories, some changing only 
one. In particular, the modified gravity model tracks showed a 
wiggle extending outside the standard model band. 
An alternate approach of convolving the expansion and growth 
observables into a single function $\fs H^{2/3}$ had the interesting 
property that it must go to a well determined constant at high 
redshift for any cosmology with a matter dominated epoch. 

The redshift ranges of particular interest are those near the wiggle, 
i.e.\ the range $z\approx0.5$--1 at which surveys will excel, but 
also $z\lesssim0.5$ for which upcoming peculiar velocity surveys such 
as TAIPAN and those using highly calibrated supernovae will offer a 
new window. Finally, the epochs at $z\gtrsim3$, perhaps accessible in 
the future through 21 cm surveys, could play a useful role in breaking 
degeneracies and carrying out a robust consistency test of early 
matter domination. Contrasting and conjoining cosmic expansion and 
growth histories provides a fundamental test of general relativity 
and insight into the intertwined evolution of spacetime as a whole 
and the large scale structure within it.

\acknowledgments 

I thank the Aspen Center for Physics, which is supported by NSF grant 
PHY-1066293, for a motivating environment. This work is supported in 
part by the Energetic Cosmos Laboratory and by 
the U.S.\ Department of Energy, Office of Science, Office of High 
Energy Physics, under Award DE-SC-0007867 and contract no.\ DE-AC02-05CH11231.


\end{document}